\documentclass{article}
\usepackage{authblk}

\usepackage{amsmath} 
\usepackage{amsfonts}	
\usepackage{amsthm} 
\usepackage{graphicx} 
\usepackage{subfigure} 
\usepackage{times}
\usepackage{helvet}
\usepackage{multirow}
\usepackage{bigstrut}
\usepackage{booktabs}
\usepackage{multirow,bigstrut}
\usepackage{tabu}
\usepackage{authblk}
\usepackage{color}
\usepackage{appendix}
\usepackage{url} 
\usepackage{enumitem}

\usepackage[sort&compress,numbers]{natbib} 

\begin{document}

\title{Followers Are Not Enough: A Question-Oriented Approach to Community Detection in Online Social Networks}

 \author[1]{David Darmon}
 \author[2]{Elisa Omodei}
 \author[3]{Joshua Garland}
 \affil[1]{University of Maryland\\
 Dept. of Mathematics}
 \affil[2]{
 LaTTiCe (CNRS, ENS, Paris 3)\\
 ISC-PIF}
 \affil[3]{
 University of Colorado at Boulder\\
 Dept. of Computer Science
 } 



\maketitle


\maketitle

\begin{abstract}

Community detection in online social networks is typically based on the analysis of the explicit connections between users, such as ``friends" on Facebook and ``followers" on Twitter. But online users often have hundreds or even thousands of such connections, and many of these connections do not correspond to real friendships or more generally to accounts that users interact with. We claim that community detection in online social networks should be question-oriented and rely on additional information beyond the simple structure of the network. The concept of `community' is very general, and different questions such as ``whom do we interact with?" and ``with whom do we share similar interests?" can lead to the discovery of different social groups. In this paper we focus on three types of communities beyond structural communities: activity-based, topic-based, and interaction-based. We analyze a Twitter dataset using three different weightings of the structural network meant to highlight these three community types, and then infer the communities associated with these weightings. We show that the communities obtained in the three weighted cases are highly different from each other, and from the communities obtained by considering only the unweighted structural network. Our results confirm that asking a precise question is an unavoidable first step in community detection in online social networks, and that different questions can lead to different insights about the network under study.
\end{abstract}

\newpage

\section{Introduction}

Networks play a central role in online social media services like Twitter, Facebook, and Google+. These services allow a user to interact with others based on the online social network they curate through a process known as contact filtering~\cite{cazabet2012automated}. For example, `friends' on Facebook represent reciprocal links for sharing information, while `followers' on Twitter allow a single user to broadcast information in a one-to-many fashion. Central to all these interactions is the fact that the \emph{structure} of the social network influences how information can be broadcast or diffuse through the service.

Because of the importance of structural networks in online social media, a large amount of work in this area has focused on using structural networks for community detection. In this traditional view, `community' is defined as a collection of nodes (users) within the network which are more highly connected to each other than to nodes (users) outside of the community 
~\cite{girvan2002a, newman2004finding}. For instance, in~\cite{java2009we}, the authors use a follower network to determine communities within Twitter, and note that conversations tend to occur within these communities. The approach of focusing on the structure of networks makes sense for `real-world' sociological experiments, where obtaining additional information about user interactions may be expensive and time-consuming. However, with the prevalence of large, rich data sets for online social networks, additional information beyond the structure alone may be incorporated, and these augmented networks more realistically reflect how users interact with each other on social media services~\cite{nguyen2011adaptive},~\cite{grabowicz2012social}.

A large body of work exists on methods for automatic detection of communities within networks, {\it e.g.,} the stochastic blockmodel~\cite{Holland1983} (and its recent generalization to weighted networks~\cite{Aicher26062014}), the Newman and Girvan algorithm~\cite{newman2004finding}, clique percolation~\cite{PalEtAl05}, Infomap~\cite{Rosvall08mapsof}, Louvain~\cite{blondel2008fast}, and the recently introduced OSLOM~\cite{LancichinettiPlos}.
All these methods \emph{begin} with a given network, and then attempt to uncover structure present in the network, i.e., they are agnostic to how the network was constructed. As opposed to this agnostic analysis, we propose and illustrate the importance of a question-focused approach. We believe that in order to understand the communities present in a data set, it is important to begin with a clear picture of the community type under consideration---{\it e.g.,} are we looking for groups of friends? or for groups of people having a common interest?---and then perform community detection with that community type in mind. 

This is especially true for social network analysis. In online social networks, a `community' could refer to several possible structures. The simplest definition of community, as we have seen, might stem from the network of explicit connections between users on a service (friends, followers, etc.). On small time scales, these connections are more or less static, and we might instead determine communities based on who is talking to whom, providing a more dynamic picture. On a more abstract level, a user might consider themselves part of a community of people discussing similar topics. We might also define communities as collections of people who exhibit similar activity profiles and seem to ``influence" each other's activity on the network. We can characterize these types of communities based on the types of questions we might ask about them:
\begin{itemize}
	\item \textbf{Structure-based:} Who are your stated friends? Whom do you follow?
	\item \textbf{Activity-based:} Whose activity influences your activity?
	\item \textbf{Topic-based:} What do you talk about?
	\item \textbf{Interaction-based:} Whom do you communicate with?
\end{itemize}

This is not meant to be an exhaustive list, but rather a list of some of the more common types of communities observed in social networks. We propose looking at when and how communities motivated by these different questions overlap, and whether different approaches to asking the question, ``What community are you in?'' leads to different insights about a social network. For example, a user on Twitter might connect mostly with computational social scientists, utilize the service nearly every time a particular user or group of users is active, talk mostly about machine learning, and interact solely with close friends (who may or may not be computational social scientists). Each of these different `profiles' of the user highlight different views of the user's social network, and represent different types of communities. We divide our approaches into four categories based on the questions outlined above: structure-based, activity-based, topic-based, and interaction-based. The structure-based approach, as outlined above, is most common, and for our data relies on reported follower relationships.

The activity-based approach is motivated by the question of ``Which user's activity `influences' the activity of another user?" 
 With this question in mind, a community can then be thought of as those individuals who ``influence" each other's activity on the network. In general, quantifying the influence of one user on another is a difficult and ill-understood problem. For that reason, we focus on a smaller subset of that problem, \emph{viz.}, quantifying influence purely from an information theoretic framework.

To accomplish this, we consider each user on Twitter as an information processing unit, but completely ignore the \emph{content} of their tweets. We then weight the directed edges (the reported follower-followee relationships) between users with the so-called \emph{transfer entropy} introduced in~\cite{schreiber2000measuring}.  Transfer entropy is an information theoretic measure of \emph{directed} information flow. It has been shown in~\cite{ver2012information} that a positive transfer entropy between a user $u$ and a follower $f$ indicates that $u$ ``influences" $f$, or that $u$ and $f$ share a common influencer. In particular, positive transfer entropy from the tweet history of a user $u$ to a follower $f$'s tweet history implies that this relationship is causal in the Granger sense~\cite{granger1963economic}. As a result, transfer entropy is often thought of as a measure of ``influence" in an information theoretic sense.  Thus, communities detected in this way have members whose tweet histories are causally related in a Granger sense, \emph{i.e.}, members who ``influence" each others activity on the network.

We want to reiterate that we do not imply that this information theoretic measure completely captures social influence in general. However, we conjecture, it does capture a useful subset of that relationship. In this paper, when we say \emph{influence} we explicitly refer to a reduction in uncertainty of whether a follower $f$ will be active or not given the past history of a user $u$. We will however show later in this paper that this information theoretic measure of influence agrees with other concepts of influence such as the Forbes list of ``Top 10 Social Media Influencers" and corroborates with the so-called strength of weak ties~\cite{granovetter1973strength}.

This activity-based approach was motivated by a method used by Shalizi \emph{et al.} in ~\cite{shalizi2007discovering} to detect functional communities within populations of neurons, but to the best of our knowledge, this is the first use of transfer entropy for community detection in online social networks. Various other information theoretic approaches have been used successfully to analyze online social networks, \emph{e.g.,} to gain insight into local user behavior~\cite{darmon2013understanding}, to detect communities based on \emph{undirected} information flow~\cite{darmon2013detecting}, and to perform network inference and link prediction~\cite{ver2012information}.

Our topic-based and interaction-based approaches, in contrast to the activity-based approach, rely on the \emph{content} of a user's interactions and ignore their temporal components. The content contains a great deal of information about communication between users.
There are two broad approaches to topic-based communities in the literature. \cite{rossi2012conversation} used a set of users collected based on their use of a single hashtag, and tracked the formation of follower and friendship links within that set of users. In~\cite{lim2012following}, the authors chose a set of topics to explore, and then seeded a network from a celebrity chosen to exemplify a particular topic. Both approaches thus begin with a particular topic in mind, and perform the data collection accordingly. Other approaches use probabilistic models for the topics and treat community membership as a latent variable~\cite{yin2012latent}.
For example, a popular approach to analyzing social media data is to use Latent Dirichlet Allocation (LDA) to infer topics based on the prevalence of words within a status~\cite{zhao2011comparing,michelson2010discovering}. The LDA model can then be used to infer distributions over latent topics, and the similarity of two users with respect to topics may be defined in terms of the distance between their associated topic distributions. Because our focus is not on topic identification, we apply a simpler approach using hashtags as a proxy for topics~\cite{becker2011beyond,tsur2012s}. We can then define the similarity of two users in terms of their hashtags, and use this similarity to build a topic-based network.

Finally, the interaction-based approach relies on the meta-data and text of messages to identify whom a user converses with on the social media service. On Twitter, we can use mentions (indicating a directed communication) and retweets (indicating endorsement of another user) to identify conversation. Moreover, we can define a directed influence between two users by considering the attention paid to that user compared to all other users. This allows us to generate a network based on conversations and user interactions.
A few works have investigated this type of community. For example, \cite{conover2011political} considered both mention and retweet networks in isolation for a collection of users chosen for their political orientation. In~\cite{deitrick2013mutually}, the authors construct a dynamic network based on simple time-windowed counts of mentions and retweets, and use the evolution of this network to aid in community detection.

Previous research on communities in social networks focused almost exclusively on different network types in isolation.
A notable exception to the analysis of isolated types of communities is~\cite{lim2012tweets}, that considered both structure-based and interaction-based communities on Twitter. However, this study focused on data collected based on particular topics (country music, tennis, and basketball), and not on a generic subpopulation of Twitter users. Moreover, it did not explore the differences in community structure resulting from the different network weightings, and focused on aggregate statistics (community size, network statistics, etc.). Another notable exception is~\cite{kao2013talison}, where the authors used a tensor representation of user data to incorporate retweet and hashtag information into a study of the social media coverage of the Occupy Movement. The tensor can then be decomposed into factors in a generalization of the singular value decomposition of a matrix, and these factors can be used to determine `salient' users. However, this approach focused on data for a particular topic (the Occupy Movement) and did not collect users based on a structural network.

The activity-based, topic-based, and interaction-based networks allow us to build a more complete picture of the \emph{implicit} networks present in online social media, as opposed to the \emph{explicit} social network indicated by structural links. In this paper, we explore the relation between these various possible networks and their corresponding communities. We begin by describing the methodologies used to generate the various types of networks, and infer their community structure. We then explore how the communities of users differ depending on the type of network used. Finally, we explore how communication patterns differ across and within the different community types.

\section{Methodology}

In the following sections, we introduce the problem of community detection, and present the data set used for our analyses. We then describe our methodology for constructing the question-specific networks. In particular, we introduce an information theoretic statistic for activity-based communities, a retweet-mention statistic for interaction-based communities, and a hashtag similarity metric for defining topic-based communities.

\subsection{Community Detection}

As discussed in the introduction, we adopt the standard definition of \emph{community}: a collection of nodes (users) within a network who are more densely connected to each other than with the rest of the network. Structural community detection is a well studied problem and several different methods and algorithms have been proposed. For a complete review of this subject we refer the reader to \cite{porter2009communities, fortunato2010community}. In this paper however we focus on a class of networks and communities that is far less studied, in particular we study networks which are both \textit{weighted} and \textit{directed} and communities within those weighted directed networks that can (but need not) \emph{overlap}. When selecting a detection algorithm we propose that all three (weight, direction, and overlap) are important for the following reasons. First, communication on Twitter occurs in a directed manner, with users broadcasting information to their followers. An undirected representation of the network would ignore this fact, and could lead to communities composed of users who do not actually share information. Second, we are interested in not just the structure of links but also in their function, and to capture this we use edge weightings which must be incorporated into the community detection process. Finally, since people can belong to multiple and possibly overlapping social (e.g., college friends, co-workers, family, etc.) and topical (e.g., a user can be interested in both cycling and politics and use the network to discuss the two topics with the two different communities) communities, we are interested in finding \textit{overlapping} communities, rather than partitions of the weighted directed network. 

This last criterion in particular poses a problem because the majority of community detection algorithms developed so far are built to find partitions of a network and few are aimed at finding overlapping communities \cite{Aicher26062014,BaumesGKMP05,PalEtAl05,ZhaWanZha07,Gre07,PhysRevE.77.016107,Lancichinetti2009,PhysRevE.80.016105,Kovacs2010}. Among these methods, even fewer deal with directed or weighted networks. For example, the work of~\cite{PalEtAl05} on clique percolation can account for both features, but not at the same time. A recent method proposed in~\cite{LancichinettiPlos}, OSLOM (Order Statistics Local Optimization Method), is one of the first methods  able to deal with all of these features simultaneously. This method of Lancichinetti et al. relies on a fitness function that measures the statistical significance of clusters with respect to random fluctuations, and attempts to optimize this fitness function across all clusters. 
Following~\cite{LancichinettiPlos}, the significance measure is defined as the probability of finding the cluster in a network without community structure. The null model used is basically the same as the one adopted by Newman and Girvan in \cite{newman2004finding} to define modularity, {\it i.e.} it is a model that generates random graphs with a given degree distribution. The authors tested their algorithm on different benchmarks (LFR and Girvan-Newman) and real networks (such as the US air transportation and the word association network), and compared its performance against the best algorithms
currently available (i.e the ones mentioned in the introduction), and found excellent results. Moreover, they showed that OSLOM is also able to recognize the absence, and not simply the presence, of community structure, by testing it on random graphs. This feature of the algorithm plays an important role in the analysis of real data networks, since it is not always the case that a community structure is indeed present and it is therefore useful to be able to detect its absence too. 
Therefore, given its versatility and performance with benchmark networks, in this paper we used OSLOM to detect \emph{overlapping} communities present in our \emph{weighted} and \emph{directed} networks.

\subsection{The Initial Dataset and Network Construction}

The dataset for this study consisted of the tweets of 15,000 Twitter users over a 9 week period (from April 25th to June 25th 2011). The users are embedded in a network collected by performing a breadth-first expansion from a random seed user. In particular, once the seed user was chosen, the network was expanded to include his/her followers, but only included users considered to be `active' (i.e., users who tweeted at least once per day over the past one hundred days). Network collection continued in this fashion by considering the active followers of the active followers of the seed, and so on until 15,000 users were added to the network. The only biasing steps in this procedure are the selection of the seed node and the filtering out of inactive users. However, because all of the community types we are interested in involve some aspect of the interaction between users, each other, and their content, the procedure provides a relatively question-agnostic follower-followee network for use with the different question-based analyses.

Since our goal is to explore the functional communities of this network, we filter the network down to the subset of users which actively interact with each other (e.g., via retweets and mentions). We do this by measuring what we call (incoming/outgoing) information events. We define an outgoing information event for a given user $u$ as either a mention made by $u$ of another user in the network, or a retweet of one of $u$'s tweets by another user in the network. The logic for this definition is as follows: if $u$ mentions a user $v$ this can be thought of as $u$ directly sending information to $v$, and if $u$ is retweeted by $v$ then $v$ received information from $u$ and rebroadcast it to their followers. In either case there was information outgoing from $u$ which affected the network in some way. Analogously, we define the incoming information event for $u$ as either being mentioned by a different user in the network, or as retweeting another user in the network.
With (incoming/outgoing) information events defined we filtered the network by eliminating all users with less than a total of 9 outgoing \emph{and} incoming information events, {\it i.e.,} less than one information event per type per week on average. 
We then further restricted our analysis to the strong giant connected component of the network built from the (incoming/outgoing) information filtered set of users. 
In this study the link is directed from the user to the follower because this is the direction in which information flows. Thus, for a pair of users $u$ and $v$, an edge $a_{v \to u}$ in the structural network has weight 1 if user $u$ follows $v$, and 0 otherwise. The final network consists of 6,917 nodes and 1,481,131 edges.

\subsection{Activity-Based Communities and Transfer Entropy}
\label{sec:method-activity}

For the activity-based communities, we consider only the timing of each user's tweets and ignore any additional content. From this starting point, we can view the behavior of a user $u$ on Twitter as a point process, where at any instant $t$ the user has either emitted a tweet ($X_{t}(u) = 1$) or remained silent ($X_{t}(u) = 0$). This is the view of a user's dynamics taken in~\cite{darmon2013understanding} and~\cite{ver2012information}. Thus, we reduce all of the information generated by a user on Twitter to a time series $\{ X_{t}(u)\}$ where $t$ ranges over the time interval for which we have data (9 weeks in this case). Because status updates are only collected in discrete, 1-second time intervals, it is natural to consider only the discrete times $t = 1 \text{s}, 2 \text{s}, \ldots, $ relative to a reference time. 

Operationally, we expect users to interact with Twitter on a human time scale, and thus the natural one-second time resolution is too fine; since most humans do not write tweets on the time scale of seconds. For this reason, we coarsen each time series by considering non-overlapping time intervals ten minutes in length. For each time interval, we record a 1 if the user has tweeted during that time interval, and a 0 if they have not. Thus, the new coarsened time series now captures whether or not the user has been active on Twitter over any given ten minute time interval in our data set. We can then compute the flow of information between two users $u$ and $v$ by computing the transfer entropy between their time series $X_{t}(u)$ and $X_{t}(v).$ See Appendix A for a detailed introduction to transfer entropy and its estimation from data.

For the communities based on transfer entropy, we weighted each edge from a user $u$ to a follower $f$ by the estimated transfer entropy of the user $u$ on $f$, 
\begin{align}
	w_{u \to f}^{\text{TE}(k)} = \widetilde{\text{TE}}_{X(u) \to X(f)}^{(k)}. \label{Eqn-EW-activity}
\end{align}
As discussed in the introduction, a positive value for the transfer entropy of the user $u$ on $f$ indicates that $u$ influences $f$, or that $u$ and $f$ share a common influencer~\cite{ver2012information}, and that the time series $X(u)$ and $X(f)$, are casually related in a Granger sense.  

 We computed the transfer entropy on each coarsened time series with lag $k$ ranging from 1 to 6, this corresponds to a lag of ten minutes to an hour. The choice of lag must balance a trade-off between additional information and sparsity of samples: as we increase the lag $k$, we account for longer range dependencies, but we also decrease the number of samples available to infer a higher dimensional predictive distribution. See Appendix A for a discussion of these issues. As we will show, the underlying communities resulting from the different lags have similar structure.

\subsection{Interaction-Based Communities and Mention / Retweet Weighting}
\label{sec:method-interaction}

Retweets and mentions are two useful features of Twitter networks which can be used to track information flow through the network.
With mentions, users are sending directed information to other users and with retweets users are rebroadcasting information from a user they follow to all of their followers. 
This type of information flow defines a community in a much different way than transfer entropy. Instead of defining communities by the loss of uncertainty in one user's tweeting history based on another's, we define interaction-based communities by weighting the user-follower network with a measure proportional to the number of mentions and/or retweets between users. 
For the interaction-based communities we consider three weighting schemes: proportional retweets,
\begin{equation}
w_{u \to f}^{\text{R}}=pR=\frac{\mbox{\# retweets of }u \mbox{ by }f}{\mbox{\# total retweets made by }f},
\end{equation}
proportional mentions,
\begin{equation}
w_{u \to f}^{\text{M}} = pM = \frac{\mbox{\# mentions of }f \mbox{ by }u}{\mbox{\# total mentions of }f},
\end{equation}
and mention-retweet as the arithmetic mean of these two measures,
\begin{align}
	w_{u \to f}^{\text{MR}} = \frac{(pM+pR)}{2}. \label{Eqn-EW-interaction}
\end{align} 


\subsection{Topic-Based Communities and Hashtag Weighting}
\label{sec:method-topic}

The final community we consider is a topic-based or topical community, i.e., a community defined by the content (topics) users discuss. In a topical community, users are defined to be a member of a community if they tweet \emph{about} similar topics as other members of the community. 
In order to detect the topical communities, we weight the edges of the user-follower network through a measure based on the number of common hashtags between pairs of users. Hashtags are a good proxy for a tweet's content as hashtags are explicitly meant to be keywords indicating the topic of the tweet. Moreover they are widely used and straightforward to detect. 

To this end, we characterize each user $u$ by a vector $\vec{h}(u)$ of length equal to the number of unique hashtags in the dataset, and whose elements are defined as
\begin{equation}
h_i(u) = \phi_i(u)\log{ \frac{N}{n_i} }
\end{equation}
where $\phi_i(u)$ is the frequency of hashtag $i$ occuring in user $u$'s tweets, $N$ is the total number of users, and $n_i$ is the number of users that have used the hashtag $i$ in their tweets. This adapted term frequency--inverse document frequency (tf-idf) measure \cite{salton_introduction_1983} captures the importance of a hashtag in the users's tweets through the first factor, but at the same time smooths it through the second factor by giving less importance to hashtags that are too widely used (as $\frac{N}{n_i}$ approaches one, its logarithm approaches zero). 

For the topical communities we weight each directed edge from a user $u$ to a follower $f$ with the cosine similarity of their respective vectors $\vec{h}(u)$ and $\vec{h}(f)$:
\begin{align}
	w_{u \to f}^{\text{HT}} = \frac{\vec{h}(u) \cdot \vec{h}(f)}{\left|\left|\vec{h}(u)\right|\right| \ \ \left|\left|\vec{h}(f)\right|\right|}. \label{Eqn-EW-topic}
\end{align}
This weight captures the similarity between users in terms of the topics discussed in their tweets. 

\section{Results and Discussion}

\subsection{Comparing Aggregate Statistics of Community Structure}

We begin by examining the overall statistics for the communities inferred by OSLOM using the weightings defined in the previous sections. The number of communities by community type is given in Table~\ref{Table-comm_count}. We see that the topic- and interaction-based networks admit the most communities. The activity-based network admits the fewest communities.  One advantage of OSLOM over many other community detection algorithms is that it explicitly accounts for singleton `communities': those nodes who do not belong to \emph{any} extant communities. This is especially important when a network is collected via a breadth-first search, as in our network, where we begin from a seed node and then branch out. Such a search, once terminated, will result in a collection of nodes on the periphery of the network that may not belong to any community in the core.


We see in Table~\ref{Table-comm_count} that the topic- and interaction-based communities have the most singletons. This result for the interaction-based community is partially an artifact of the retweet/mention weighting: 717 of the users were disconnected from the network by how the weights were defined, resulting in `orphan' nodes which we have included in the collection of singletons for all of our analyses. However, even after accounting for this artifact, the interaction-based network still has the most non-orphan singletons. This seems to indicate that a large fraction of the 6917 (nearly 25\%) do not interact with each other in a concerted way that would mark them as a community under our interaction-based definition. This agrees with a result previously reported in~\cite{romero2011influence} about how most users passively interact with incoming information on Twitter.

\begin{table}[ht]
	\caption{Number of non-singleton communities and singletons by community type: S(tructural), A(ctivity-based), T(opic-based), and I(nteraction-based).}
	\centering
	\begin{tabular}{| c | c | c |}
		\hline Community Type & \# of Communities & \# of Singletons \\ \hline
		S & 201 & 308 \\
		A, Lag 1 & 101 & 951 \\
		A, Lag 2 & 99 & 600 \\
		A, Lag 3 & 106 & 611 \\
		A, Lag 4 & 105 & 668 \\
		A, Lag 5 & 107 & 632 \\
		A, Lag 6 & 106 & 642 \\
		T & 289 & 1064 \\
		I & 252 & 2436 (1719) \\ \hline
	\end{tabular}
	\label{Table-comm_count}
\end{table}

\begin{figure}[!htbp]
  \centering
\includegraphics[width=\textwidth]{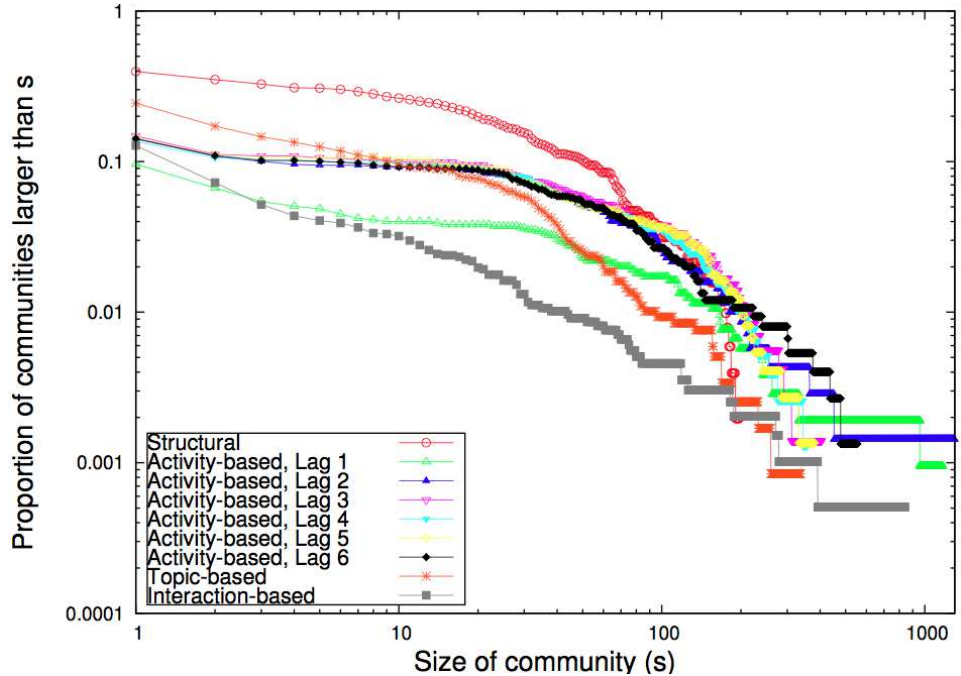}
\caption{The proportion of communities greater than $s$ in size, across the different community types. Note the logarithmic scale on the horizontal and vertical axes.}
\label{Fig-community_size_distribution}
\end{figure}

Next we consider the distribution of community sizes across the community types. The complementary cumulative distribution of community sizes is given in Figure~\ref{Fig-community_size_distribution}. Note that both axes are plotted on log-scales. Thus, for a fixed community size $s$, Figure~\ref{Fig-community_size_distribution} shows the proportion of communities of size greater than $s$ for each community type. We see that the community distributions have longer tails for the non-structural networks, and that the interaction-based network has the longest tail.
Note that the activity-based communities, using transfer entropy estimated with varying lags, seem to converge on a similar large-scale community structure around lag 3. That is, the communities based on lag 1 and lag 2 transfer entropies are generally larger, and these communities resolve into smaller communities as the lag increases beyond 2. The distribution of community sizes for lag larger than 2 is insensitive to the lag.
Most importantly, we see that the distributions of community sizes differ across the community types, highlighting that the different networks give rise to different large-scale community structure dependent on the particular weighting of the structural network.

\subsection{Comparing Community Structure with Normalized Mutual Information}

In the previous section, we saw that the large scale statistics of the communities were highly dependent on the type of community under consideration. However, macroscale network statistics do not account for differences in community structure that result from operations such as splitting or merging of communities. Moreover, this view does not account for which users belong to which communities, and in particular which users belong to the same communities across community types. To answer this question, we invoke methods for the comparison of clusters: given two different clusterings of nodes into communities, how similar are the two clusters? The standard approach to answering this question is to define a metric on the space of possible partitions.  
Because we detect coverings rather than partitions, standard cluster comparison metrics like variation of information~\cite{meilua2003comparing} are not appropriate. Instead, we use a generalization of variation of information first introduced in~\cite{Lancichinetti2009}, the normalized mutual information. The normalized mutual information stems from treating clustering as a community identification problem: given that we know a node's community membership(s) in the first covering, how much information do we have about its community membership(s) in the second covering, and vice versa? Consider the two coverings $\mathcal{C}_{1}$ and $\mathcal{C}_{2}.$ We think of the community memberships of a randomly chosen node in $\mathcal{C}_{1}$ as a binary random vector $\mathbf{X} \in \{0, 1\}^{|\mathcal{C}_{1}|}$ where the $i^{\text{th}}$ entry of the vector is 1 if the node belongs to community $i$ and 0 otherwise. Similarly, $\mathbf{Y} \in \{ 0, 1\}^{|\mathcal{C}_{2}|}$ is a binary random vector indicating the community memberships of the node in $\mathcal{C}_{2}$. Then the normalized mutual information is defined as
\begin{align}
	\text{NMI}(\mathcal{C}_{1}, \mathcal{C}_{2}) = 1 - \frac{1}{2} \left( \frac{H[\mathbf{X} | \mathbf{Y}]}{H[\mathbf{X}]} + \frac{H[\mathbf{Y} | \mathbf{X}]}{H[\mathbf{Y}]}\right)
\end{align}
where $H[\cdot]$ denotes a marginal entropy and $H[\cdot | \cdot]$ denotes a conditional entropy. The normalized mutual information varies from 0 to 1, attaining the value of 1 only when $\mathcal{C}_{1}$ and $\mathcal{C}_{2}$ are identical coverings up to a permutation of their labels. See the appendix of~\cite{Lancichinetti2009} for more details.

\begin{figure}[!htbp]
  \centering
  \begin{tabu}{cl}
    \toprule
    Normalized Mutual Information & Community Types \\
    \midrule
    \multirow{8}[8]{*}{\includegraphics[width=0.50\textwidth]{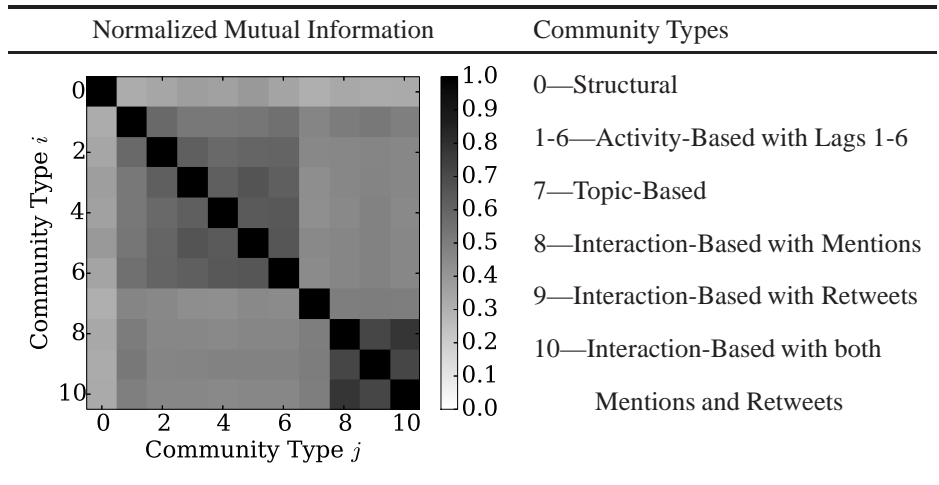}}&0---Structural \bigstrut\\
    &1-6---Activity-Based with Lags 1-6\bigstrut\\
    &7---Topic-Based \bigstrut\\
    &8---Interaction-Based with Mentions\bigstrut\\
    &9---Interaction-Based with Retweets\bigstrut\\
    &10---Interaction-Based with both \bigstrut\\
    &\,\,\,\,\,\,\,\,\,\,\,\,\,\,Mentions and Retweets\bigstrut\\
    &\bigstrut\\
    \bottomrule
\end{tabu}
\caption{The normalized mutual information between the coverings inferred from the different community types. 
Values of normalized mutual information close to 1 indicate similarity in the community structure, while values close to 0 indicate dissimilarity. The normalized mutual information is computed with singletons and orphan nodes included.}
\label{Fig-compare_coverings}
\end{figure}

We considered the normalized mutual information between the communities inferred from the structural network and the networks weighted with lag 1 through 6 transfer entropies, hashtag similarity, and mention, retweet, and mention-retweet activity. The resulting $\text{NMI}(\mathcal{C}_{i}, \mathcal{C}_{j})$ are shown in Figure~\ref{Fig-compare_coverings}. We see that similarity between the coverings is dictated by the generic community type (structural, activity-based, etc.). That is, the transfer entropy coverings are more similar to each other than to any of the other coverings, with a similar result for the mention, retweet, and mention-retweet coverings. Interestingly, the coverings resulting from the different weightings are all more similar to each other than to the structural covering from the unweighted network. Also note that the covering based on the hashtag similarities are different from all of the other weight-based coverings. In agreement with the results reported in the previous section, we see that the activity-based communities share similar structure for lags greater than 2. Because of these two results, in the remainder of the paper, we focus on the activity-based communities inferred using the lag-4 transfer entropy.

Thus, we see that although the activity-based, interaction-based, and topic-based communities relied on the structural network, their community structure differs \emph{the most} from the community structure of the follower network. This agrees with the results from the previous section, and reinforces that the follower network is a necessary but not sufficient part of detecting communities characterized by properties beyond follower-followee relationships.

\subsection{Comparing Edges Across Different Community Types}

\begin{figure}[!htbp]
	\centering
	\includegraphics[width=0.8\textwidth]{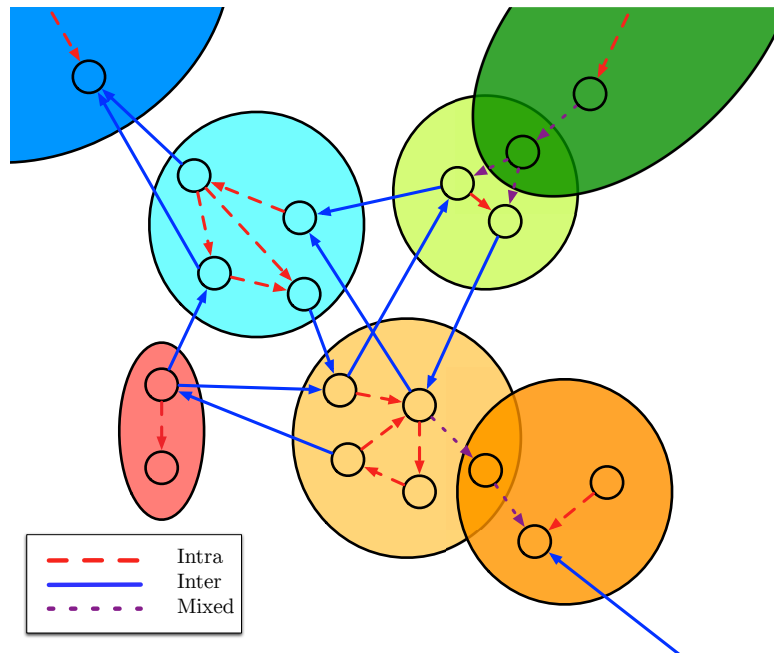}
	\caption{A schematic of how, given a covering, the edges of the network can be partitioned using (\ref{Eqn-edge_types}) into inter-edges, intra-edges, and mixed-edges. Inter-edges (red dashed) cross community boundaries. Intra-edges (blue solid) remain inside community boundaries. Mixed-edges (purple dotted) both remain in and cross community boundaries due to overlap in community membership. Each column corresponds to the same collection of weights, but partitioned using a different covering. Each row corresponds to the same covering, but for the different weights.}
	\label{Fig-edge_types}
\end{figure}

Any covering determined by OSLOM induces a natural partition of the edges in a directed network. In particular, let $u$ and $f$ be two users in the network, and let $M_{u}$ and $M_{f}$ be their community memberships. Then any edge $e_{u \to f}$ can be partitioned into one of three classes by
\begin{align}
	T(e_{u \to f}) = \left\{ \begin{array}{ll}
		\text{Inter-edge} &: \ \ \ M_{u} \cap M_{f} = \emptyset \\
		\text{Intra-edge} &: \ \ \ M_{u} = M_{f}\\
		\text{Mixed-edge} &: \ \ \ \text{ otherwise} \\
	\end{array}\right. \label{Eqn-edge_types}.
\end{align}
In words, an inter-edge connects two users who share no community memberships, an intra-edge connects two users who each belong to the same communities, and a mixed-edge connects two users who share some, but not all, community memberships. Thus, inter-edges cross community boundaries, intra-edges lie within community boundaries, and mixed-edges lie at the borders of community boundaries. See Figure~\ref{Fig-edge_types} for a schematic of this edge partitioning.

\begin{figure}[!h]
	\centering
	\includegraphics[width=1\textwidth]{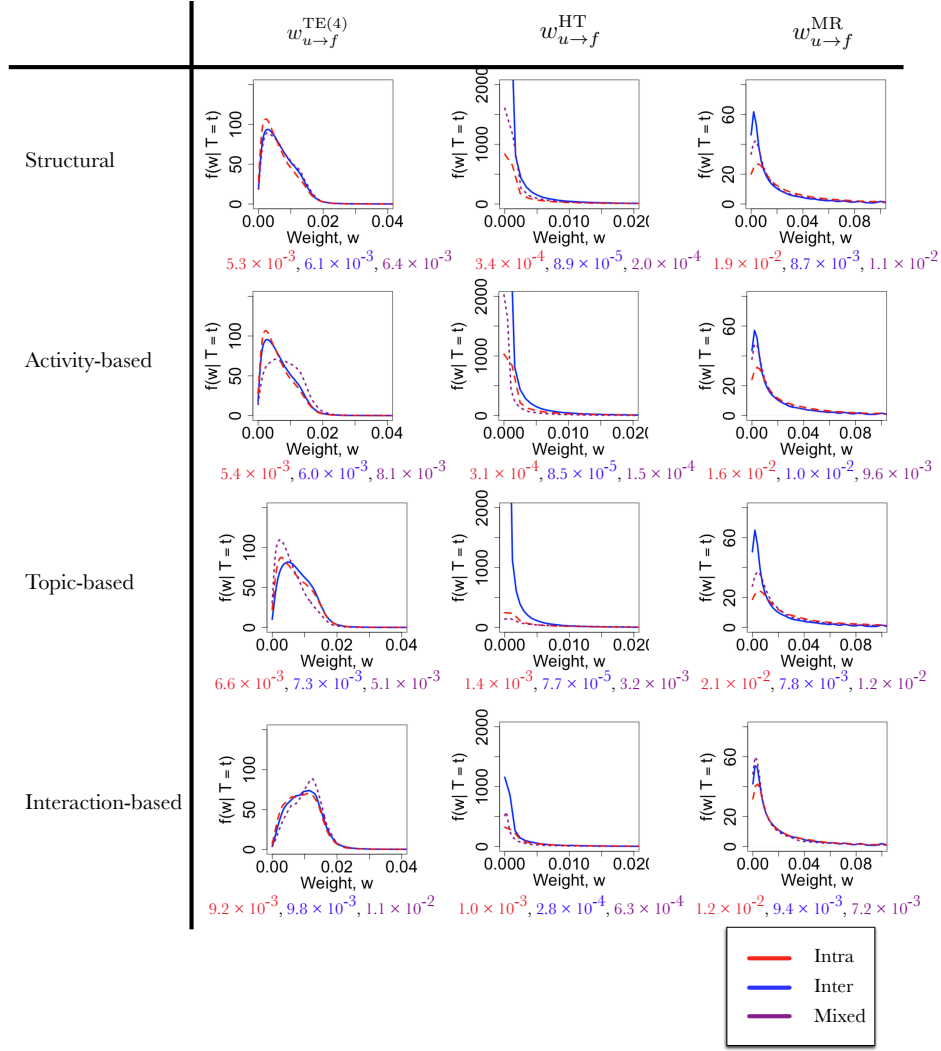}
	\caption{The density of edge weights for different community types (rows) and weight types (columns). The red,  blue, and purple values below each collection of densities indicate the median of weight on intra-, inter-, and mixed-edges, respectively.}
	\label{Fig-distributions_by_types}
\end{figure}

Each community type (Structural, Activity-based, Topic-based, Interaction-based) induces a different partition of the edges, and each edge type (Transfer Entropy, Hashtag, Mention-Retweet) induces a different distribution of weights. That is, let $W_{u \to f}$ be the weight on an edge $E_{u \to f}$ chosen uniformly at random from the network, and compute
\begin{align}
	P(W_{u \to f} \leq w_{u \to f} | T(E_{u \to f}) = t),
\end{align}
the empirical distribution over the edge weights conditioned on the edge type $t$ being one of inter, intra, and mixed. For meaningful community structure, we expect the distribution of edge weights to differ with the edge type $t$. If the community structure were arbitrary, the edge weights would be independent of the edge type, and all three conditional distributions would be identical.
The densities associated with these distributions are shown in Figure~\ref{Fig-distributions_by_types}.

Each \emph{column} of Figure~\ref{Fig-distributions_by_types} demonstrates how the density of weights change with the community type used to induce the partition. For example, the second column shows how the density of hashtag weights change based on the community partition. In general, the densities differ in non-trivial ways across the edge types and coverings. However, we see that the medians of the distributions provide a useful summary statistic. Under each collection of density functions, we list the median value of the inter, intra, and mixed-edges in red, blue, and purple, respectively. Unsurprisingly, we see that the greatest difference between the distributions occurs when we consider matching Community-type / Weight-type pairs, since the communities are determined based on the corresponding weights. However, we also see differences in the distributions when the Community-type / Weight-type pairs do not match. For example, for the covering corresponding to the activity-based communities, the hashtag weights tend to be higher for edges internal to communities than between communities. Similarly, for the covering corresponding to the topic-based communities, the transfer entropy weights tend to be higher on edges within (intra) and between (inter) communities, and lower for edges between members with multiple, non-identical memberships (mixed).

We also note that the distribution of transfer entropy weights always tend to be higher for edges crossing community boundaries (inter- or mixed-edges) compared to those edges within community boundaries (intra-edges). This is a property not exhibited by either the hashtag or mention-retweet weights, independent of the covering used to partition the edges. Moreover, for all but the activity-based covering, the weights of mixed-edges tend to be highest overall.  Recall that the transfer entropy $\text{TE}_{X(u) \to X(f)}$ quantifies the reduction in uncertainty about a follower $f$'s activity from knowing the activity of a user $u$. This result therefore implies that, in terms of prediction, it is more useful to know the time series of a user followed outside of the community compared to a user followed inside of the community, and even more useful to know the time series of a user that shares some, but not all, of the same community memberships. Thus, in an information theoretic sense, we see that novel information useful for prediction is more likely to flow \emph{across} community boundaries than \emph{within} community boundaries.

\subsection{Qualitative Analysis of Community Memberships Across Types}

As demonstrated by~\cite{good2010performance} in the context of modularity maximization-based community detection, an exponential number of nearby partitions may exist that nearly maximize an objective function used to measure the goodness-of-fit of a graph partition used for community detection. Because of this and related issues, it is always wise to perform some sort of qualitative study of the communities returned by any  community detection algorithm to verify their meaningfulness with respect to the scientific question at hand. In this section, we consider a collection of communities in such a study. 

In the topic-based communities, we find a single community consisting of 83 users who tweet about environmental issues and frequently use hashtags such as \#green, \#eco and \#sustainability. We also find a different community of 47 users who tweet about small businesses and entrepreneurship, using hashtags such as \#smallbiz, \#marketing and \#enterpreneur. In both cases most members of the topic-based communities are not found in the same community in the other networks, indicating that while these people talk about the same things and can therefore be considered a community based on their content, they do not strongly interact with each other nor behave the same, and so belong to different social groups with respect to interactions and behavior.

Another interesting example is a community whose topics tend to focus on Denver and Colorado. These users do not belong to the same community in the interaction-based network, but most of them do belong to the same community in the activity-based network. This indicates that these users react to the same events and issues regarding Colorado and are therefore strongly connected in the topic-based and activity-based networks, but at the same time they do not directly interact with each other and are therefore more loosely connected in the interaction-based networks, where they belong to different communities.

As an aside, it is interesting to examine who the most influential users are in these communities, as determined by transfer entropy. Recall that influence in this context refers to a reduction in uncertainty about a follower(s) activity on Twitter given the tweet history of a followee. Said differently, higher transfer entropy, or a higher reduction in uncertainty, implies more influence. Interestingly, among the most influential users in these communities, those with highest transfer entropy on average, we find \@Colorado, which is the state official Twitter account, \@ConnectColorado, a page created to connect Coloradans, and the CBS Denver account, a popular news agency. This means that these accounts had high influence in terms of their followers' activity on Twitter. This is not surprising as these communities discuss Colorado and these accounts provide information on this topic. However, this does provide a check that this measure of activity-based influence makes sense and suggests this could be a useful measure for influence on social media if used with other measures. This is discussed more in future work. 

Lastly, it is interesting to note that two of the Forbes ``Top 10 Social Media Influencers", 
 happen to be in our network, \emph{viz.}, Ann Tran and Jessica Northey, and transfer entropy also quantified these two users as highly influential.
This suggests that activity-based or information-theoretic influence may be useful in finding influential members purely based on their temporal tweet history while completely ignoring both tweet content and social status.
\section{Conclusion and Future Work}



In this study, we have demonstrated that the communities observed in online social networks are highly question-dependent. The questions posed about a network \emph{a priori} have a strong impact on the communities observed.  Moreover, using different definitions of community reveal different and interesting relationships between users. More importantly, we have shown that these different views of the network are not revealed by using the structural network or any one weighting scheme alone. By varying the questions we asked about the network and then deriving weighting schema to answer each question, we found that community structure differed across community types on both the macro (e.g. number of communities and their size distribution) and micro (e.g. specific memberships, comemberships) scale in interesting ways.


To verify the validity of these communities we demonstrated that boundaries between communities represent meaningful internal/external divisions. In particular, conversations (e.g. retweets and mentions) and topics (e.g. hashtags) tended to be most highly concentrated within communities. We found this to be the case even when the communities were defined by a different criterion from the edge weights under study. 



At first glance the boundaries defined by the activity-based communities derived from the transfer entropy weighting seemed less meaningful. However, upon further investigation our novel use of transfer entropy for the detection of activity-based communities highlighted an important fact about this social network: influence tended to be higher across community boundaries than within them. This result echos the `strength of weak ties' theory from~\cite{granovetter1973strength}, which has found empirical support in online social networks~\cite{grabowicz2012social}. This means that our use of transfer entropy not only defines boundaries that are meaningful divisions between communities but also illustrates that users who have a strong influence on a community need not be a member of that community.

Our findings may have important implications to a common problem in social network analysis: identification of influential individuals---but further work must be carried out. Many network measures of influence are based on the various types of centrality (degree, betweenness, closeness, eigenvector, etc.)~\cite{newman2009networks}. Most centralities depend explicitly on the structure of the network under consideration. 
But we have seen in our study that a structural network alone is not sufficient to capture user interaction or even activity-based influence in online social media. Thus, a na\"ive application of centrality measures to a structural network for influence detection may give rise to erroneous results. This result has been explored previously~\cite{kitsak2010identification}, and our work further highlights its importance. 

Based on our preliminary results, we conjecture that weighted generalizations of these centrality measures using transfer entropy might lead to better insights about who is actually influential in an online social network. In addition to exploring this phenomenon further, we plan to explore a broader selection of choices for both the transfer-entropy lag and tweet history time resolution. We believe that by doing an in-depth analysis of both of these parameters we can discover interesting activity-based communities that occur on much broader time scales.

This work demonstrates that asking the proper question, and then crafting an appropriate weighting scheme to answer that question, is an unavoidable first step for community detection in online social media. More generally, this work illustrates that without a clear definition of community, many rich and interesting communities present in online social networks remain invisible. Question-oriented community detection can bring those hidden communities into the light.

\section{Funding}
This work was partially supported by the R\'{e}gion \^{I}le-de-France [DIM 2011 to E.O.]; and by the French National Agency of Research [ANR-12-CORD-0018 to E.O.].

\section{Acknowledgments}

Thanks to Cesar Flores, Lu\'is Seoane, Kevin Stadler, Jody Wright, and Nix Barnett for their contributions to preliminary ideas related to this paper and to Michelle Girvan, William Rand, Aaron Clauset, and Ryan James for their valuable comments and suggestions. Finally, we would like to acknowledge the Santa Fe Institute and its Complex Systems Summer School for providing the intellectually stimulating environment where this project began.

\appendix

\section{Appendix A: Transfer Entropy and Its Estimation from Data}

Let $\{X_{t}\}$ and $\{ Y_{t}\}$ be two strong-sense stationary stochastic processes\footnote{Recall that a stochastic process is strong-sense stationary if the joint distribution for the process evaluated at finitely many time points is invariant to an overall timeshift~\cite{grimmett2001probability}.}. In our work, these would correspond to the activities, $X_{t}(u)$ and $X_{t}(v),$ of two users $u$ and $v$. We use the notation $X_{t-k}^{t}$ to denote the values of the stochastic process from time $t-k$ to time $t$, $X_{t-k}^{t} = (X_{t-k}, X_{t-(k-1)}, \ldots, X_{t - 1}, X_{t})$. The lag-$k$ transfer entropy~\cite{schreiber2000measuring} of $Y$ on $X$ is defined as 
\begin{align}
	\text{TE}_{Y \to X}^{(k)} &= H\left[X_{t} | X_{t-k}^{t-1}\right] - H\left[X_{t} | X_{t-k}^{t-1}, Y_{t-k}^{t-1}\right], \label{Eqn-TE}
\end{align}
where
\begin{align}
	H\left[X_{t} | X_{t-k}^{t-1}\right] = - E\left[ \log_{2} p(X_{t} | X_{t-k}^{t-1}) \right]
\end{align}
and 
\begin{align}
	H\left[X_{t} | X_{t-k}^{t-1}, Y_{t-k}^{t-1}\right] = - E\left[ \log_{2} p(X_{t} | X_{t-k}^{t-1}, Y_{t-k}^{t-1}) \right]
\end{align}
are the usual conditional entropies over the conditional (predictive) distributions $p(x_{t} | x_{t-k}^{t-1})$ and $p(x_{t} | x_{t-k}^{t-1}, y_{t-k}^{t-1})$. This formulation was originally developed in~\cite{schreiber2000measuring}, where transfer entropy was proposed as an information theoretic measure of \emph{directed} information flow. Formally, recalling that $H\left[X_{t} | X_{t-k}^{t-1}\right]$ is the uncertainty in $X_{t}$ given its values at the previous $k$ time points, and that $H\left[X_{t} | X_{t-k}^{t-1}, Y_{t-k}^{t-1}\right]$ is the uncertainty in $X_{t}$ given the joint process $\{(X_{t}, Y_{t})\}$ at the previous $k$ time points, transfer entropy measures the reduction in uncertainty of $X_{t}$ by including information about $Y_{t-k}^{t-1},$ controlling for the information in $X_{t - k}^{t-1}$. By the `conditioning reduces entropy' result~\cite{cover2012elements}
\begin{align}
	H[X | Y, Z] \leq H[X | Y],
\end{align}
we can see that transfer entropy is always non-negative, and is zero precisely when $$H\left[X_{t} | X_{t-k}^{t-1}\right] = H\left[X_{t} | X_{t-k}^{t-1}, Y_{t-k}^{t-1}\right]$$, in which case knowing the past $k$ lags of $Y_{t}$ does not reduce the uncertainty in $X_{t}$. If the transfer entropy is positive, then $\{ Y_{t}\}$ is considered causal for $\{ X_{t}\}$ in the Granger sense~\cite{granger1963economic}.

When estimating transfer entropy from finite data, we will assume that the process $\{(X_{t}, Y_{t})\}$ is jointly stationary, which gives us that
\begin{align}
	p(x_{t} | x_{t-k}^{t-1}) = p(x_{k+1} | x_{1}^{k})
\end{align}
and
\begin{align}
	p(x_{t} | x_{t-k}^{t-1}, y_{t-k}^{t-1}) = p(x_{k+1} | x_{1}^{k}, y_{1}^{k})
\end{align}
for all $t$. That is, the predictive distribution only depends on the past, not on when the past is observed. Given this assumption, we compute estimators for $p(x_{k+1} | x_{1}^{k})$ and $p(x_{k+1} | x_{1}^{k}, y_{1}^{k})$ by `counting': for each possible marginal and joint past $x_{1}^{k}$ and $(x_{1}^{k}, y_{1}^{k})$, we count the number of times a future of type $x_{k+1}$ occurs, and normalize to obtain the appropriate estimators of the one-step-ahead predictive distributions. Call these estimators $\hat{p}(x_{k+1} | x_{1}^{k})$ and $\hat{p}(x_{k+1} | x_{1}^{k}, y_{1}^{k})$. Then the plug-in estimator for the transfer entropy is
\begin{align}
	\widehat{\text{TE}}_{Y \to X}^{(k)} &= \hat{H}\left[X_{t} | X_{t-k}^{t-1}\right] - \hat{H}\left[X_{t} | X_{t-k}^{t-1}, Y_{t-k}^{t-1}\right]
\end{align}
where we use the plug-in estimators $\hat{H}\left[X_{t} | X_{t-k}^{t-1}\right]$ and $\hat{H}\left[X_{t} | X_{t-k}^{t-1}, Y_{t-k}^{t-1}\right]$ for the entropies. It is well known that the plug-in estimator for entropy is biased~\cite{paninski2003estimation}. To account for this bias, we use the Miller-Madow adjustment to the plug-in estimator~\cite{miller1955note}. For a random variable $X$ taking on finitely many values from an alphabet $\mathcal{X}$, the Miller-Madow estimator is
	\begin{align}
		\tilde{H}[X] = \hat{H}[X] + \frac{|\hat{\mathcal{X}}| - 1}{2 n}
	\end{align}
	where $|\mathcal{\hat{X}}|$ is the number of observed symbols from the alphabet $\mathcal{X}$ and $n$ was the number of samples used to estimate $\hat{H}[X].$ The definition of transfer entropy~(\ref{Eqn-TE}) can be rewritten in terms of joint entropies as
	\begin{align}
		TE_{Y \to X}^{(k)} &= H[X_t | X_{t-k}^{t-1}] - H[X_t | X_{t-k}^{t-1},Y_{t-k}^{t-1}] \\ 
		&= H[X_t,X_{t-k}^{t-1}]-H[X_{t-k}^{t-1}]-H[X_t,X_{t-k}^{t-1},Y_{t-k}^{t-1}]+H[X_{t-k}^{t-1},Y_{t-k}^{t-1}],
	\end{align}
	We then apply the Miller-Madow adjustment individually to each of the entropy terms. For example, for the first term, we have
	\begin{align}
		\tilde{H}[X_{t}, X_{t - k}^{t-1}] = \tilde{H}[X_{t - k}^{t}] = \hat{H}[X_{t-k}^{t}] + \frac{|\widehat{\mathcal{X}^{k+1}}| - 1}{2n},
	\end{align}
	where $|\widehat{\mathcal{X}^{k+1}}|$ is the number of $(k + 1)$-tuples we actually observe (of the $2^{k + 1}$ possible tuples). Doing this for each term, the overall Miller-Madow estimator for the transfer entropy is
	\begin{align}
		\widetilde{TE}_{Y \to X}^{(k)} &= \tilde{H}[X_t | X_{t-k}^{t-1}] - \tilde{H}[X_t | X_{t-k}^{t-1},Y_{t-k}^{t-1}] \\ 
		&= \tilde{H}[X_t,X_{t-k}^{t-1}]-\tilde{H}[X_{t-k}^{t-1}]-\tilde{H}[X_t,X_{t-k}^{t-1},Y_{t-k}^{t-1}]+\tilde{H}[X_{t-k}^{t-1},Y_{t-k}^{t-1}].
	\end{align}
	One possible problem with this estimator is that it can result in \emph{negative} estimates of entropies. That usually occurs when $\hat{H}$ is very small. In these cases, we truncate set the estimator to zero.


\bibliographystyle{comnet}
\bibliography{references}

\end{document}